\documentclass[a4paper,twoside]{article}
\newcommand{\affil}[1]{$^{\rm #1}$}
\usepackage[authoryear]{natbib}
\bibpunct{(}{)}{;}{a}{}{,}
\usepackage{graphicx}
\usepackage{amssymb}  
\date{} 
\newcommand{\Msun}{{\rm M}_{\odot}}

\newcommand{\iso}[2]{\hbox{${}^{#1}{\rm #2}$}}

\title{\large\bf\flushleft  Heavy element abundances in planetary 
nebulae: A theorist's perspective}
\author{\parbox{\textwidth}{\flushleft
\vspace{-0.5cm}
{\it Amanda I. Karakas\affil{A,C} and Maria Lugaro\affil{B}}\\
\vspace{0.4cm}
{\small \affil{A}\,Research School of Astronomy \& Astrophysics,
Mount Stromlo Observatory, Cotter Road Weston Creek, ACT 2611, Australia}\\
{\small \affil{B}\,Centre for Stellar and Planetary Astrophysics, 
Monash University, PO Box 28M, Clayton VIC 3800, Australia}\\
{\small \affil{C}\,Email: akarakas@mso.anu.edu.au}}}
\begin{document}
\twocolumn[
\begin{changemargin}{.8cm}{.5cm}
\begin{minipage}{.9\textwidth}
\vspace{-1cm}
\maketitle
%
%
\small{\bf Abstract:} The determination of heavy element
abundances from planetary nebula (PN) spectra provides an exciting 
opportunity to study the nucleosynthesis occurring in the progenitor 
asymptotic giant branch (AGB) star.  We perform post-processing
calculations on AGB models of a large range of mass and 
metallicity to obtain predictions for the production of 
neutron-capture elements up to the first $s$-process peak at 
strontium. We find that solar metallicity intermediate-mass AGB 
models provide a reasonable match to the heavy element composition 
of Type I PNe. Likewise, many of the Se and Kr enriched PNe
are well fitted by lower mass models with solar or close-to-solar
metallicities. However the most Kr-enriched objects, and the PN
with sub-solar Se/O ratios are difficult to explain with AGB
nucleosynthesis models. Furthermore, we compute $s$-process 
abundance predictions for low-mass AGB models of very low metallicity 
(Fe/H $\approx -2.3$) using both scaled solar and an
$\alpha$-enhanced initial composition. For these models, O is 
dredged to the surface, which means that abundance ratios measured 
relative to this element (e.g., X/O) do not provide a reliable
measure of initial abundance ratios, or of production within the
star owing to internal nucleosynthesis.


\medskip
\medskip
\end{minipage}
\end{changemargin}
]
\small

\section{Introduction}

After the thermally-pulsing AGB (TP-AGB) phase is terminated, low to
intermediate-mass stars ($\sim 0.8$ to $8\Msun$) evolve at near constant
luminosity to become post-AGB objects. If the ejected envelope has
sufficient time to become ionized by the hot central star before
dissipating, then the object will also be observed as a planetary 
nebula (PN). For recent reviews of TP-AGB and post-AGB stars, see
\citet{herwig05} and \citet{vanwinckel03}, respectively. 
The ionised nebula is comprised of material from the convective
envelope that once surrounded the core, hence, nebular abundances
may reveal information about the efficiency of mixing events and
chemical processing that took place during previous evolutionary phases,
in addition to the initial composition of the parent star
\citep[e.g.,][]{dopita97,magrini09}. 

\begin{figure}
\begin{center}
\includegraphics[scale=0.35,angle=0]{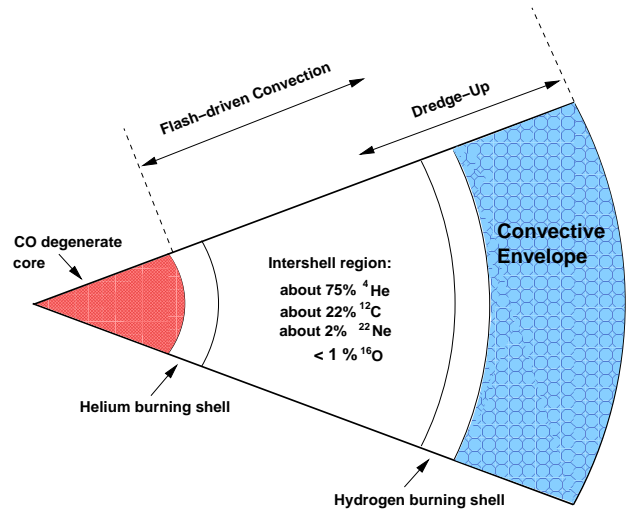}\\
\caption{Schematic structure of an AGB star.
\label{agb}}
\end{center}
\end{figure}

During the life of a star, mixing between the nuclear-processed core
and envelope may occur, and this has the effect of enriching the
surface composition in the products of H-burning (first and second
dredge-up; hot bottom burning), and He-burning (the third dredge-up).
Note that the first dredge-up occurs after core H-exhaustion, whereas
the second dredge-up after core He-exhaustion. Both the third dredge
up (TDU) and hot bottom burning (HBB) occur during the TP-AGB phase.
We refer the reader to, e.g., \citet{boothroyd99} for details 
on the first and second dredge up.

A TP-AGB star is characterised by two nuclear burning shells
above a degenerate C-O core, surrounded by a deep convective envelope
(see Fig.~\ref{agb}). The TDU mixes material from the 
He-intershell to the envelope. This material is composed primarily
of \iso{4}He and \iso{12}C, along with trace quantities of 
heavy elements produced by the $slow$ neutron capture process 
(the $s$-process).
The TDU may occur after each instability of the helium shell,
slowly enriching the envelope in carbon and heavy elements.  HBB
occurs in intermediate-mass AGB stars ($\sim 4$ to 8$\Msun$) when 
the base of the convective envelope can dip into the top of the 
hydrogen burning shell. 

Abundances derived from PN spectra provide a complimentary data 
set to the abundances derived from the spectra of cool evolved stars.
For example, the elemental abundances of He, Ne, and Ar can be 
obtained, along with the abundances of C, N, O, S, and Cl 
\citep{aller83,stanghellini00,leisy06}. Recent observations have
also revealed that some PNe are enriched in heavy elements that 
can be produced by the $s$-process including Ge, Se, Kr, Xe, and
Ba \citep[][hereafter SD08]{sharpee07,sterling08}. SD08 obtained 
Se and Kr abundances for 120 Galactic PNe, and investigated trends
between $s$-process enrichments and PN morphology and other
nebular and stellar characteristics. It was found that while 
some PN have large enrichments of Se and Kr (e.g., [Se,Kr/O] $> 1$),
Type I PN have lower $s$-process abundances, on average, than the
sample as a whole. 

In \citet{karakas09} we compared nucleosynthesis predictions from 
models of intermediate-mass AGB stars to the results from SD08.
Here we briefly summarize the results from Karakas et al. (2009), and 
provide new results for lower mass
AGB models of solar and halo compositions.

\section{Type I Planetary Nebula}

The aims of \cite{karakas09} were to compare the 
$s$-process predictions from intermediate-mass AGB stars to 
the Se and Kr compositions of the Type I PN identified in the 
SD08 sample. Such a comparison may be used to constrain the
efficiency of the TDU in AGB models, which is not well determined. 
Also this information may help elucidate to what extent 
intermediate-mass AGB stars contribute to the Galactic 
inventory of $s$-process elements.
The motivation to use Type I PN comes from their high He/H and 
N/O ratios characteristic of proton-capture nucleosynthesis 
\citep{stanghellini06}, as well as spatial distributions and 
kinematics indicative of a young population \citep{corradi95}
with initial masses between $\approx 2.5$ to $6\Msun$.

Furthermore, we had also hoped to identify the neutron 
source(s) operating in intermediate-mass AGB stars. In low-mass 
stars it has been shown that the \iso{13}C($\alpha,n$)\iso{16}O 
reaction is the dominant source of neutrons, whereas the 
\iso{22}Ne($\alpha,n$)\iso{25}Mg reaction only plays a minor role 
\citep{lambert95,gallino98}
The \iso{22}Ne($\alpha,n$) reaction has been suggested as the 
dominant neutron source in intermediate-mass AGB 
stars \citep{truran77,garcia06}.    The $s$-process abundance 
pattern that results from each reaction is quite different, owing
to differing time-scales over which the neutrons are released
along with maximum neutron densities that vary by orders of 
magnitude: $\sim 10^{8}$ for the \iso{13}C reaction compared to 
$\sim 10^{13}$ neutrons per cm$^{-3}$ for \iso{22}Ne. One 
important result, for example, is the prediction of low Rb/Sr ratios
in low-mass AGB stars, which has been observationally verified
\citep[e.g.,][]{lambert95}.

The details of the stellar models and nucleosynthesis calculations
have been discussed in detail in \citet{karakas09} and \citet{karakas07b}.
Here we summarize the main findings. It is important to point 
out that only models 
with HBB ($>4\Msun$) would be observed as Type I PNe, whereas 
observations put the minimum mass at $\sim 2.5\Msun$. Clearly, the 
lowest mass range would require either rapid stellar rotation,
extra mixing, and/or the effect of a binary companion to assist 
in producing the Type I composition.

\begin{figure}
\begin{center}
\includegraphics[scale=0.3,angle=270]{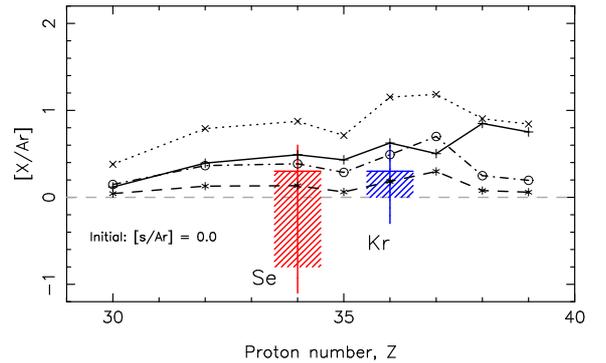}\\
\caption{\small{The surface abundances of neutron-capture elements
around the first $s$-process peak. The solid line indicates
abundances from the 3$\Msun$, $Z = 0.02$ model, the dashed line
abundances from the 6.5$\Msun$, $Z = 0.02$ model, the 
dot-dashed line abundances from the 5$\Msun$, $Z = 0.008$,
and the dotted line abundances from the 5$\Msun$, $Z = 0.004$
model. Abundances are taken at the tip of the TP-AGB, after
the last computed thermal pulse. 
The boxes represent the range
of observed Se and Kr abundances for Type I PNe, see the
text for details.}
\label{fig2}}
\end{center}
\end{figure}

In \citet{karakas09} we found that the HBB AGB models of solar 
or near solar metallicity (in particular $Z= 0.02$ and 0.008) 
were a reasonable match to the Se and Kr abundance distribution
of the Type I PNe, see Fig.~\ref{fig2}. In Fig.~\ref{fig2} 
the abundances are shown as [X/Ar] ratios, and are plotted as a 
function of proton number, $Z$. The hatched boxes around the 
elements Se and Kr indicate the range of the Type I PN abundances
from the SD08 sample, with the solid vertical line extending the
boxes by an extra 0.3~dex, indicative of the typical errors.

The intermediate-mass low metallicity models ($Z = 0.004$) 
produced Se and Kr abundances that were in excess of the 
Type I PN abundances (see the dotted line in Fig.~\ref{fig2}), 
indicating that the PN 
we see today did not form from such low-metallicity objects. 
This result is consistent with the PN evolving from a young, 
disk population. The 3$\Msun$ models of solar metallicity were 
a marginal fit to the Se and Kr data for Type I PNe.  
The 3$\Msun$ models became carbon rich and we would not 
predict them to be observed as Type I objects with high He/H 
and N/O ratios. Rapid rotation during the main sequence could, 
however, bring some CN-processed material to the surface.

Binarity may be the condition necessary for shaping bipolar 
PNe (and perhaps all PNe) \citep{moe06}. Many, but not all, 
Type I PNe 
are also bipolar in shape.  If binary interactions can produce
a Type I composition, then this may lead to the TP-AGB phase
being terminated early, before many thermal pulses have occurred.
Hence the prediction would be for lower $s$-process elements
to be produced, on average, compared to star that evolve to the
tip of the AGB phase. This would also be consistent with
the observations of no $s$-process enrichment observed in 
Type I PNe.

\section{Se and Kr abundances in low-mass AGB stars}

\begin{figure}
\begin{center}
\includegraphics[scale=0.3,angle=270]{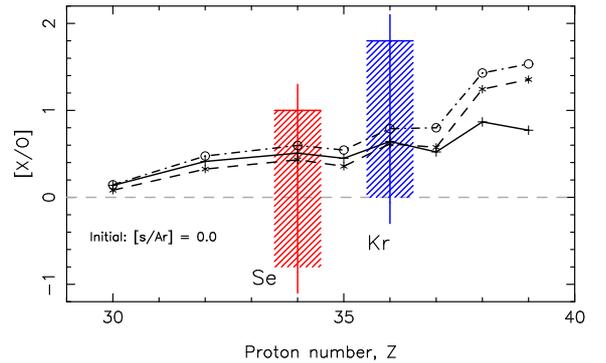}\\
\caption{\small{The surface abundances of neutron-capture elements
around the first $s$-process peak. The solid line indicates
abundances for the 3$\Msun$, $Z = 0.02$ model, the dashed line
abundances for the 2.1$\Msun$, $Z = 0.008$ model, and the 
dot-dashed line abundances for the 2.5$\Msun$, $Z = 0.008$
model. Abundances are taken at the tip of the TP-AGB, after
the last computed thermal pulse. The boxes represent the range
of observed Se and Kr abundances for the full PN sample, see 
text for details.}
\label{fig3}}
\end{center}
\end{figure}

In Fig.~\ref{fig3} we show the surface abundances of Zn,
Ge, Se, Br, Kr, and Sr at the tip of the AGB phase for three
low-mass AGB models. The solid line represents the abundances
from the 3$\Msun$, $Z = 0.02$ model, the dashed line abundances
for the 2.1$\Msun$, $Z = 0.008$ model, and the dot-dashed line
abundances for the 2.5$\Msun$, $Z = 0.008$ model. Abundances
are shown as [X/O] ratios. The hatched boxes around the elements 
Se and Kr indicate the range of the PN abundances for the 
full sample from SD08, where the solid vertical line extending the
boxes by an extra 0.3~dex, indicative of the typical errors.
All models include a partial mixing zone
of 0.002$\Msun$. This mixing zone produces a  \iso{13}C pocket
during the interpulse period, allowing the reaction 
\iso{13}C($\alpha,n$)\iso{16}O to release free neutrons. Note 
that models without partial mixing zones result in no enrichment
of $s$-process elements \citep[e.g.,][]{karakas07a}. We refer
the reader to \citet{busso99} and \citet{herwig05} for 
detailed discussion about the formation of \iso{13}C pockets and
related uncertainties.

From Fig.~\ref{fig3} we see that the solar metallicity and $Z=0.008$
models are within the observed range of Se and Kr PN abundances. 
Starting with a scaled solar abundance distribution, with
[$s$/O] $=0$, it is not possible to obtain sub-solar Se abundances, i.e.,
[Se/O] $<$ 0. Likewise, the most Kr-enriched objects, with [Kr/O] $> 1$, 
are also not matched by the models. 

The boxes indicating the range of derived PN abundances are somewhat 
misleading, because they do not convey the Se and Kr abundance 
distribution of PN, and most importantly, the number with extreme 
Se and Kr abundances.
Fig.~4 from SD08 shows that there are only six PN out of 120 
with low Se abundances, that is [Se/O,Ar] $< -0.3$.
There are also only six PN with high Kr abundances, that is 
[Kr/O,Ar] $> 1$. Indeed, the rest of the sample of 120 PNe shows 
[Se/O,Ar] between 0 and 1, consistent with the
stellar models. Likewise, most of the Kr-enriched PN
have $0 \lesssim [{\rm Kr/O,Ar}] \lesssim 1.0$, consistent with
the AGB models plotted in Fig.~\ref{fig3}. There
are still large uncertainties in deriving the Kr and Se
abundances from PNe spectra. Hence it is possible that the most
extreme PN abundances are effected by these uncertainties.

Finally, one other possibility is that the PN with the lowest
Se abundances do not reflect internal nucleosynthesis
during the progenitor AGB phase, but the initial Se abundance.
Hence the very low Se/O abundances are the result of a peculiar
galactic chemical evolution. 
If this is the case, then it would be intriguing to see if 
chemical evolution models would be able to explain the large spread
in Se found in the full sample of PNe. 

\section{Low-metallicity PN}

\begin{figure}
\begin{center}
\includegraphics[scale=0.3,angle=270]{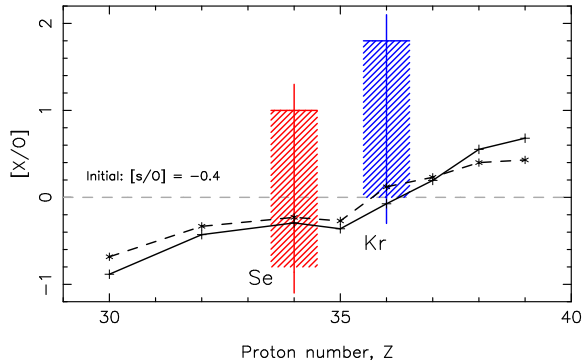}\\
\caption{\small{Same as Fig.~\ref{fig2}, except the models shown 
are the 1.25$\Msun$, [Fe/H]$=-2.3$ model (solid line), and
the 2$\Msun$, [Fe/H]$=-2.3$ model (dashed line). An $\alpha$-enhanced
initial abundance pattern was used (e.g., [O/Fe] = $+0.4$), and
scaled solar for elements heavier than iron. }
\label{fig4}}
\end{center}
\end{figure}

There have been a few low-metallicity PN identified including
BoBn 1 \citep{otsuka08,zijlstra06}.
These PN likely originate from low-mass progenitors with
initial masses less than $\sim 1\Msun$. These PN could be used
as tracers of the low-metallicity environment out of which
the progenitor stars formed, and they could be used to constrain
the evolution and nucleosynthesis of low-metallicity AGB stars.

In Fig.~\ref{fig4} we show the $s$-process abundance distribution
for two low-metallicity, low-mass AGB models. For these models we
choose an initial $\alpha$-enhanced abundance pattern (e.g.,
[O/Fe] $=+0.4$), which results in initial heavy element abundances 
with [X/O] $=-0.4$ when measured relative to oxygen. Note that 
relative to iron, the initial abundances for these elements are scaled
solar, i.e., [X/Fe] $=0$.  In Fig.~\ref{fig4} we show the observed
Se and Kr abundance distribution from the SD08 sample, noting that
most of the PN used were Galactic PNe with higher metallicities
than used in the AGB models presented here. 
From Fig.~\ref{fig4} we see that the relative ratio of Zn/O is 
sub-solar, with [Zn/O] $\sim - 1$ (starting at [Zn/O] $=-0.4$)
from both models.  If we examine the oxygen abundance, we find
that a significant amount of this element is dredged
to the surface, that is, [O/Fe] $= 0.87$ for the 1.25$\Msun$ model,
and [O/Fe]$=1.20$ for the 2$\Msun$ case. 
In both cases, Zn is actually produced by the $s$-process but 
in smaller quantities to oxygen, hence the [X/O] ratios do 
not reflect the degree of production. For example, the final 
[Zn/Fe] $=0.45$ for both models. Note that in both models, the 
iron abundance remains essentially unchanged.

Likewise for Kr, the abundances measured relative to iron are
[Kr/Fe] = 1.0~dex and 1.1~dex for the 1.25$\Msun$
and 2$\Msun$ models, respectively.  This is compared to 
[Kr/O] $\approx 0$ (see Fig.~\ref{fig4}). Hence our results indicate
that for the lowest metallicity PNe, the dredge-up of oxygen
implies that this element is no longer a suitable proxy for
the initial metallicity of the progenitor star.
Argon is sometimes used in place of 
oxygen \citep[][SD08]{leisy06}, under the assumption that 
it remains unchanged by AGB nucleosynthesis. In \citet{karakas09}
we showed that this is indeed the case. Zinc is also a potentially 
useful reference element because it does not condense into dust
as easily as iron \citep[e.g.,][]{welty99}, and can be observed in
PNe \citep{dinerstein01}. Zn is at the beginning of the $s$-process
chain, and our results show that it can be produced in 
observable quantities in low-metallicity
AGB stars. The best indicator of the metallicity is iron, 
but the element iron abundance is difficult to accurately 
determine from PN spectra \citep{perinotto99,sterling05a,delgado09}.

\section{Concluding Remarks}

Abundances derived from PN spectra are an invaluable tool to study
the late phases of stellar evolution of low and intermediate-mass stars,
as well as trace the chemical evolution of galaxies. Elemental 
abundances for C, N, O, S, the noble gases He, Ne, and Ar, as well
as elements produced by neutron capture processes 
including Ge, Se, Kr, and
Ba can be obtained, with varying degrees of accuracy. 
Observations of neutron capture elements in PNe can be used to
constrain the amount of third dredge up mixing following the final
thermal pulses of the progenitor star. In \citet{karakas09} we
attempted to use the Se and Kr abundance information for Type I
PNe, along with information on lighter element abundances (e.g.,
C, N, O, Ne etc) to constrain the neutron source operating in
intermediate-mass AGB stars. We were unable to reach firm conclusions
using the Se and Kr abundances. Certainly, it seems that the \iso{22}Ne
source is required to produce enrichments in $s$-process elements
in these more massive AGB stars. It is less clear if the \iso{13}C
neutron source is also required. Increases in the Kr abundances
were observed in models with a \iso{13}C pocket, but not beyond
the amounts observed in Type I PN spectra. Obtaining abundances
for more heavy elements beyond the first $s$-process peak may help
to distinguish among the possibilities.

Type I PNe may also originate from binary interactions, and/or by
rapid rotation during the progenitor's main sequence phase. Certainly
obtaining more abundances for Type I PNe may help to distinguish between
the single or binary evolutionary channels, and to constrain the
initial progenitor masses of these systems. 

We compared results from lower mass AGB models to the full sample of 
PNe from SD08. Our results indicate that AGB models of solar or 
near solar metallicity produce a reasonable match to Se and Kr
enrichments seen in the full sample. Our models cannot explain the
PNe with sub-solar Se abundances, nor the PNe with the largest
Kr enrichments. These outliers are only a small percentage of the
full sample ($\sim 5$\%), and may reflect either uncertainties in 
the abundance determinations or the effects of inhomogeneous 
chemical evolution. 

Finally, we present results for low-mass, low metallicity AGB
models with metallicities appropriate for the Halo ([Fe/H] $\approx -2.3$).
These AGB models were computed with an $\alpha$-enhanced initial 
composition, and scaled solar for elements heavier than iron.
In these low-metallicity AGB models, oxygen is dredged-up to the
surface, which means that measuring the enrichment of heavy elements
relative to this element (e.g., [Zn/O]) produces misleading
results. For example, [Zn/O] $\approx -1$ for these models, when
[Zn/Fe] $\approx 0.45$. Hence, it is important to determine the
level of O dredge-up occurring, or to use another, more suitable
proxy for the metallicity (e.g., Ar or Fe if possible).

\section*{Acknowledgments} 

The authors would like to thank Quentin Parker for organising 
the MASH workshop, and for his patience in waiting for this paper!
AIK acknowledges support from the Australian Research Council's
Discovery Projects funding scheme (project number DP0664105).
ML is supported by the Monash Research Fellowship.



\end{document}